# Experts' cognition-driven safe noisy labels learning for precise segmentation of residual tumor in breast cancer

Yongquan Yang[1,2], Jie Chen[1], Yani Wei[1,3], Mohammad Alobaidi[4], Hong Bu[1,3]

1. Institute of Clinical Pathology, West China Hospital, Sichuan University, Chengdu, China
2. Zhongjiu Flash Medical Technology Co., Ltd., Mianyang, China
3. Department of Pathology, West China Hospital, Sichuan University, Chengdu, China
4. Department of Civil Engineering, McGill University, Montreal, Canada

**Abstract**

Precise segmentation of residual tumor in breast cancer (PSRTBC) after neoadjuvant chemotherapy is a fundamental key technique in the treatment process of breast cancer. However, achieving PSRTBC is still a challenge, since the breast cancer tissue and tumor cells commonly have complex and varied morphological changes after neoadjuvant chemotherapy, which inevitably increases the difficulty to produce a predictive model that has good generalization with usual supervised learning (SL). To alleviate this situation, in this paper, we propose an experts' cognition-driven safe noisy labels learning (ECDSNLL) approach. In the concept of safe noisy labels learning, which is a typical type of safe weakly supervised learning, ECDSNLL is constructed by integrating the pathology experts' cognition about identifying residual tumor in breast cancer and the artificial intelligence experts' cognition about data modeling with provided data basis. Experimental results show that, compared with usual SL, ECDSNLL can significantly improve the lower bound of a number of UNet variants with 2.42% and 4.1% respectively in recall and fIoU for PSRTBC, while being able to achieve improvements in mean value and upper bound as well.

Email addresses: Yongquan Yang (remy_yang@foxmail.com), Jie Chen (jzcjedu@foxmail.com), Yani Wei (yani_wei@stu.scu.edu.cn), Mohammad H. Alobaidi (mohammad.alobaidi@mcgill.ca), Hong Bu (hongbu@scu.edu.cn).

## Introduction

Residual tumor in breast cancer (RTBC) indicates the tumor that still remains in breast cancer tissue after neoadjuvant chemotherapy, which is an important iatrotechnique in the breast cancer treatment process (Asaoka et al., 2020; Charfare et al., 2005; Mieog et al., 2007; Schott & Hayes, 2012). Commonly, RTBC is associated with invasive ductal carcinoma in which tumor has spread into surrounding breast tissue. Quantitative evaluation of RTBC can provide clues important to prognosis and subsequent therapy of breast cancer (Pu et al., 2020; Yau et al., 2022). The key point of quantitative evaluation of RTBC is to achieve precise segmentation of RTBC (PSRTBC), which is a fundamental key technique in the treatment process of breast cancer, such as be leveraged to calculate the tumor-stroma ratio that has been proven to be a prognostic factor in breast cancer (de Kruijf et al., 2011).

Whole sliding imaging (WSI) (Hanna et al., 2020), which was previously referred to as virtual microscopy, involves scanning a pathology glass slide into digital image at high resolution and displaying the digitalized image on a computer screen (Gilbertson & Yagi, 2005; Pantanowitz et al., 2018). WSI has provided the foundation for the development of digital pathology(Al-Janabi et al., 2012; Unternaehrer et al., 2020), which enables machine learning techniques, such as deep learning (DL) (LeCun et al., 2015) (mostly deep neural networks (He et al., 2016; G. Huang et al., 2017; Sandler et al., 2018; Szegedy et al., 2017; Tan & Le, 2021; Xie et al., 2017; Zoph et al., 2018)), to be carried out for various medical image-based evaluations. In recent years, based on WSI, a number of DL-based image semantic segmentation solutions have been proposed for precise segmentation of tumor in breast cancer (PSTBC) of pre-treatment biopsy (Bhattacharjee et al., 2022; Diao et al., 2022a; Elmannai et al., 2021; Priego-Torres et al., 2020, 2022). However, as the breast cancer tissue and tumor cells commonly have complex and varied morphological changes after neoadjuvant chemotherapy (Rajan et al., 2004), which makes achieving PSRTBC more difficult than achieving PSTBC. The fact that the task of PSRTBC is more difficult than the task of PSTBC has be qualitatively confirmed by pathology experts in their daily diagnosis process. Meanwhile, our recent work (Yang et al., 2024) has also qualitatively proven this fact, as the generalization of the predictive model for PSRTBC was worse than the generalization of the predictive model for PSTBC. As a result, achieving PSRTBC is still a challenge demanding prompt solution.

To alleviate this situation, in this paper, we propose an experts' cognition-driven (ECD) safe noisy labels learning (SNLL) (ECDSNLL) approach, which integrates the pathology experts' cognition and the artificial intelligence experts' cognition in a SNLL paradigm to construct an automatic system for PSRTBC. SNLL concerns about the situation where both noisy data and noisy-free data are leveraged to evolve a predictive model (Li et al., 2021). The proposed ECDSNLL approach has three novel innovations: 1) Proposing problem-conscious labeling techniques that take into account the pathological experts' cognition about identifying RTBC, which results in a large amount of noisy data (LAND) and a small amount of noisy-free data (SANFD) to more effectively describe RTBC; 2) Establishing a two-stage cascade learning paradigm (TSCLP) that takes into account the artificial intelligence experts' cognition about data modeling, which first guarantees the high recall of the target based on a noisy data (ND) and then promotes the precision of the target based on a noisy free data (NFD) to effectively achieve better performance in prediction corresponding to the target; 3) Constructing a new approach naturally in the concept of SNLL to more effectively achieve PSRTBC, by integrating the pathological experts' cognition-driven LAND and SANFD, and the artificial intelligence experts' cognition-driven TSCLP based on the data basis of ND and NFD. Regarding to its three novel innovations, the proposed ECDSNLL approach is different from the existing work (Li et al., 2021) that realized SNLL in the ensemble learning paradigm (Dietterich, 2000; Yang et al., 2023; Z. H. Zhou, 2009, 2012) and the related works that realized DL-based predictive models based on separate noisy-free data (Bhattacharjee et al., 2022; Priego-Torres et al., 2020, 2022) or separate noisy data (Diao et al., 2022b; G. Xu et al., 2019; Y. Xu et al., 2014; Yang et al., 2024).

That the pathological experts' cognition is referred to produce the data basis of LAND and SANFD is because pathological experts know how to effectively describe RTBC. That the artificial intelligence experts' cognition is referred to produce the TSCLP learning paradigm with the provided data basis of ND and NFD is because artificial intelligence experts know how to effectively achieve better performance in data modeling. That the pathological experts' cognition-driven LAND and SANFD and the artificial intelligence experts' cognition-driven TSCLP based on the data basis of ND and NFD are integrated to construct the ECDNSLL approach for PSRTBC is because LAND and SANFD naturally fit the data basis of ND and NFD required by TSCLP in the concept of SNLL. More detailed explanations about the three innovations of ECDSNLL can be found in the first subsection of section 3.

Besides our previous exploration on achieving PSRTBC via noisy label learning with diverse noisy samples (Yang et al., 2024), as far as we know, this paper is the first to particularly address the challenge of PSRTBC via safe noisy label learning with both noisy and noisy-free samples. Apart from the primary contribution which is the proposed ECDSNLL approach for PSRTBC, this paper has following additional contributions: Implementation and evaluation of ECDSNLL for PSRTBC with samples of whole slide imaging; Production of a predictive model for PSRTBC with samples of whole slide imaging; Testing the domain generalization of the produced predictive model for PSRTBC under microscope scenario.

## 2. Literature Review

In this section, we briefly review safe noisy labels learning and alternative DL-based approaches for PSTBC that are closely related to the concept of experts' cognition-driven safe noisy labels learning proposed in this paper.

### 2.1. Safe noisy labels learning

Safe noisy labels learning (SNLL) (Li et al., 2021) has a close relation to noisy labels learning (NLL)(Natarajan et al., 2013; Song et al., 2023). Different from the paradigm of NLL that is only based on noisy data, the concept of SNLL is based on both noisy data and noisy-free data, and to ensure the learning from both noisy data and noisy-free data will not be inferior to simply learning from the noisy-free data. SNLL is a typical type of safe weakly supervised learning (SWSL) (Li, 2021; Li et al., 2021), which aims to ensure that the extra weakly supervised data will not be inferior to a simple supervised learning model (Li, 2021). In the current literature, SNLL is usually realized in the ensemble learning paradigm (Dietterich, 2000; Yang et al., 2023; Z. H. Zhou, 2009, 2012). Different from the current literature, in this paper, we propose a novel experts' cognition-driven SNLL approach for PSRTBC via taking into account the pathological experts' cognition about identifying RTBC and the artificial intelligence experts' cognition about data modeling.

### 2.2 Alternative DL-based approaches for PSTBC

Existing alternative DL-based approaches for PSTBC can be classified into two schemes: 1) learning with noisy-free/accurate labels (Bhattacharjee et al., 2022; Priego-Torres et al., 2020, 2022), and 2) learning with noisy/inaccurate labels (Diao et al., 2022b; G. Xu et al., 2019; Y. Xu et al., 2014; Yang et al., 2024). The first type of scheme adopts the fully supervised learning paradigm, which is simple and clear in the scheme design. However, because the identification of RTBC is very difficult even for pathologists in some cases, large amount of accurately labeled (noisy-free) data is often rare, which will inevitably limit the generalization of the predictive model. The second type of scheme adopts the weakly supervised learning paradigm (Z.-H. Zhou, 2018), and produces the predictive model based on inaccurately labeled (noisy) data (Karimi et al., 2020; Song et al., 2022). This, to some extent, solves the problem of the difficulty in obtaining large amount of accurately labeled (noisy-free) data faced by the first type of scheme, since it is much easier to obtain large amount of inaccurately labeled (noisy) data. However, identically, because the identification of RTBC is very difficult even for pathologists in some cases, the inaccurately labeled (noisy) data often contain very complex noise, which inevitably affects the performance of the prediction model. As a result, theoretically, existing alternative DL-based approaches for PSTBC are probably not the best choices for PSRTBC and more advanced approach is needed. Different from these alternative approaches, in this paper, we propose to address the PSTBC task with an approach in the concept of SNLL which is based on both noisy data and noisy-free data.

## 3. Research Methodology

This section is structured as follows. In section 3.1, we present the methodology of ECDSNLL for PSRTBC. Based on the presented methodology of ECDSNLL for PSRTBC, we present the details for the implementation of ECDSNLL for PSRTBC in section 3.2. In section 3.3, we present the metrics, strategies and details for evaluating ECDSNLL for PSRTBC. In section 3.4, we describe the method for producing the predictive model for PSRTBC. Finally, in section 3.5, we describe the method for assessing the predictive model for PSRTBC under microscope scenario.

### 3.1. Methodology of ECDSNLL for PSRTBC

The outline of the proposed ECDSNLL approach for PSRTBC is shown as Fig. 1. More details of ECDSNLL are provided in the rest of this subsection.

Figure 1

**Outline of the proposed ECDSNLL approach for PSRTBC. ECDSNLL constitutes of the pathology experts' cognition-driven LAND and SANFD, the artificial intelligence experts' cognition driven TSCLP, and the feeding relations between them. LAND is short for large amount of noisy data; SANFD is short for small amount of noisy-free data. TSCLP is short for two-stage cascade learning paradigm which is based on deep neural networks (DNN)**

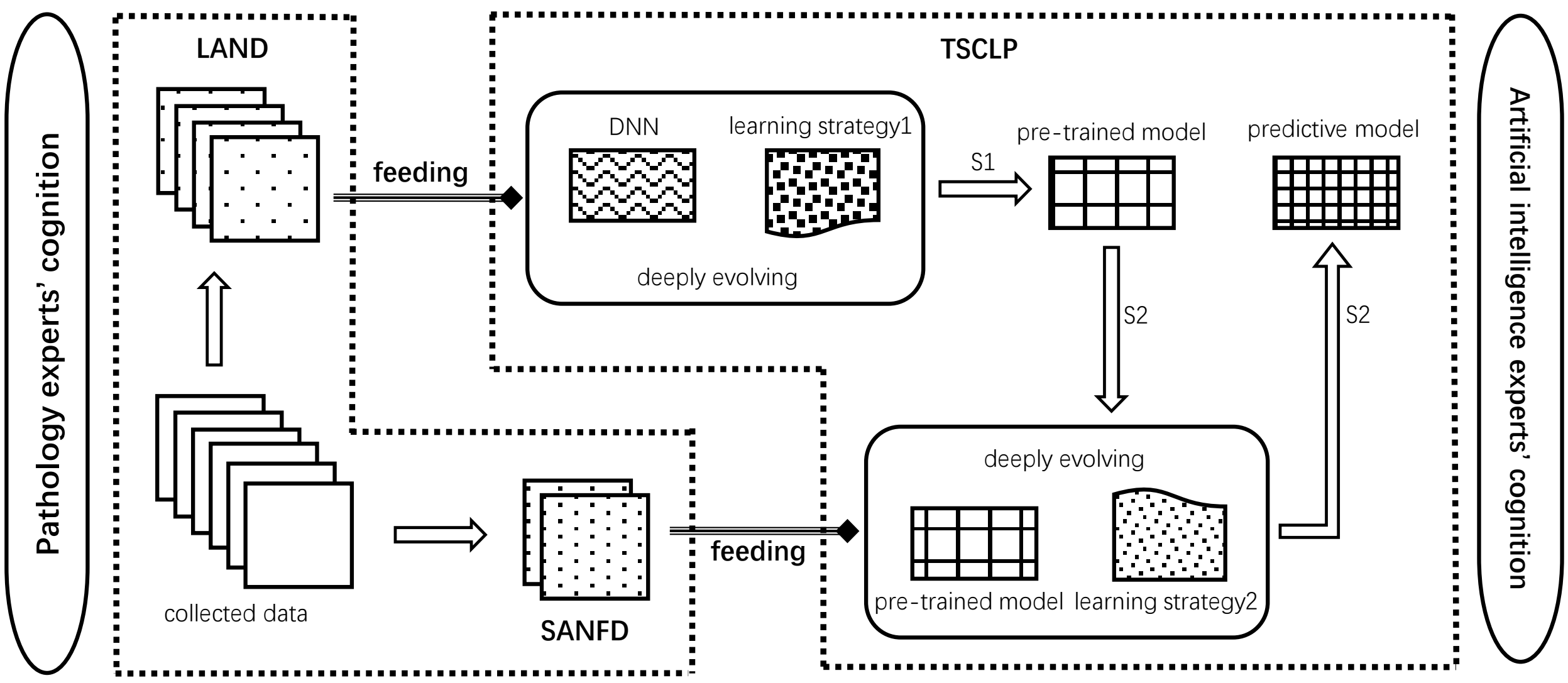

### 3.1.1. Pathology experts' cognition-driven LAND and SANFD

Referring to the pathological experts' cognition that identifying none RTBC is easier than identifying RTBC, we made two labeling rules, which resulted in a large amount of noisy data (LAND) and a small amount of noisy-free data (SANFD). Since identifying none RTBC is easier, we made labeling rules one that pathological experts can try best to exclude the none RTBC, which will only ensure the RTBC to be included in the target as much as possible to produce the LAND that may contain many none RTBC areas as the target. Including RTBC in the target as much as possible, LAND at least has a very high recall rate of RTBC. Meanwhile, since identifying RTBC is more difficult, we made labeling rule two that pathological experts can only ensure RTBC to be accurately included in the target in certain cases to produce the SANFD. Including RTBC in the target accurately, SANFD at least has a high precision rate of RTBC. As LAND and SANFD are produced by referring to the pathological experts' cognition about the identification of RTBC, the produced LAND and SANFD are pathology experts' cognition-driven.

### 3.1.2. Artificial intelligence experts' cognition-driven TSCLP

Referring to the well-known artificial intelligence experts' cognition that first guaranteeing the high recall of the target and then promoting the precision of the target can achieve better overall performance corresponding to the target (Viola & Jones, 2001, 2004), we established a two-stage cascade learning paradigm (TSCLP). TSCLP is appropriate to evolve a predictive model for the precise segmentation of the target by trenching the respective advantages of both noisy data (ND) and noisy free data (NFD). In the first stage of TSCLP, a deep neural network can be trained to produce a pre-trained model based on ND. If ND can at least have a very high recall rate of the target, the pre-trained model can achieve a high recall rate of the target in prediction. In the second stage of TSCLP, the pre-trained model can be further optimized to produce the final predictive model based on NFD. If NFD can at least have a high precision rate of the target, the final predictive model can achieve a higher precision rate of the target in prediction compared with the pre-trained model. Therefore, evolving the predictive model for the precise segmentation of the target based on ND and NFD, TSCLP can better balance the recall and precision rates of the target for prediction. As TSCLP are formed by referring to the artificial intelligence experts' cognition about achieving better overall performance in prediction, the formed TSCLP is artificial intelligence experts' cognition-driven.

#### 3.1.3. Construction of ECDSNLL for PSRTBC

Based on the analysis of LAND and SANFD in section 2.1.1, and the analysis of TSCLP in section 2.1.2, we constructed ECDSNLL for PSRTBC by feeding LAND as the base ND for the stage one (S1) of TSCLP and feeding SANFD as the base NFD for the stage two (S2) of TSCLP, as LAND and SANFD fit the data basis of ND and NFD required by TSCLP. As LAND and SANFD are pathological experts' cognition-driven, and TSCLP is artificial intelligence experts' cognition-driven, the constructed ECDSNLL is an experts' cognition-driven (ECD) approach. Meanwhile, evolving the predictive model based on the noisy data LAND and the noisy-free data SANFD, the constructed ECDSNLL is naturally a safe noisy labels learning (SNLL) approach.

### 3.2. Implementation of ECDSNLL for PSRTBC

#### 3.2.1. Preparation of LAND and SANFD

We collected 291 while slide images (WSIs) digitalized from pathology glass slides of patients after treated with neoadjuvant chemotherapy. As the coarse annotations of LAND are much easier than the accurate annotations of SANFD according to their labeling rules, we randomly selected 231 WSIs for coarse annotations to produced LAND and 60 WSIs for accurate annotations to produce SANFD. To generate LAND, we cropped 43752 tiles and corresponding labels with 128X128 pixels from the coarsely annotated areas of the selected 231 WSIs at the 5X resolution. To generate SANFD, we cropped 1865 tiles and corresponding labels with 128×128 pixels from the accurately annotated areas of the selected 60 WSIs at the 5X resolution. Some example tiles of LAND and SANFD are shown as Fig. 2.

In preparing the LAND and SANFD, any personal identifiers, such as names and dates of birth, are removed from the images through a process known as de-identification, ensuring that the data cannot be traced back to individual patients.

**Figure 2**

**Example tiles of LAND and SANFD. Top row: tiles and corresponding labels of LAND; Bottom row: tiles and corresponding labels of SANFD. The green polygon areas indicate the residual tumor labels**

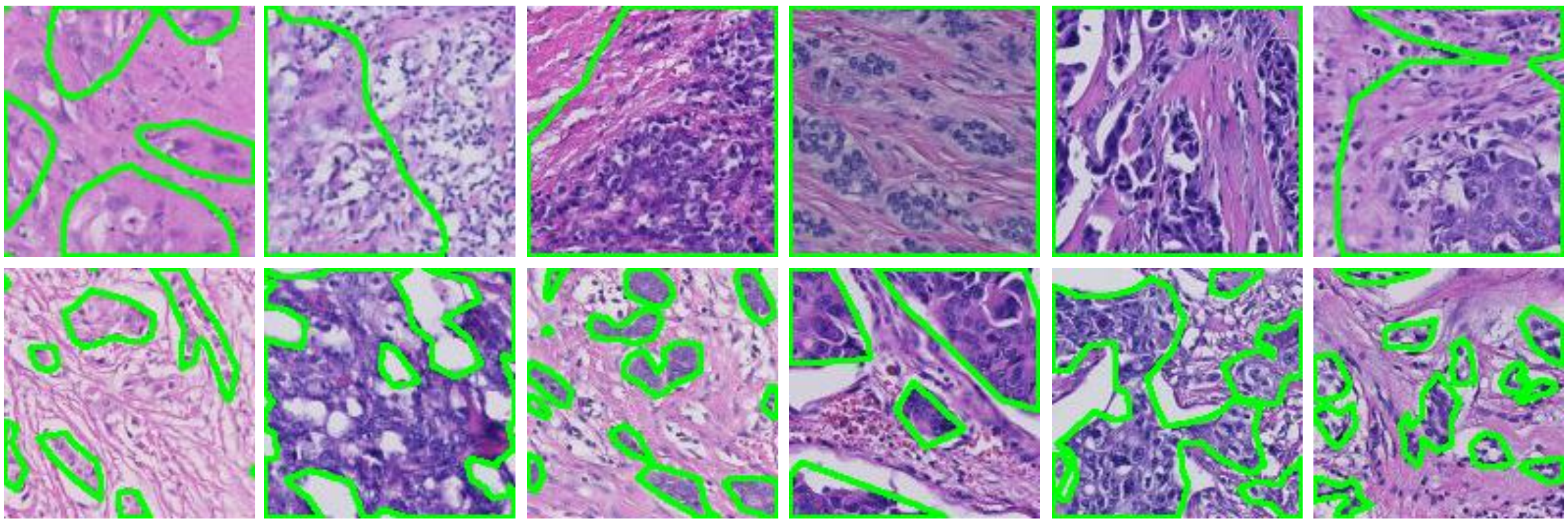

#### 3.2.2. Implementation details of TSCLP

The primary key point of the implementation of TSCLP is to employ an appropriate DNN for the predictive model. Since PSRTBC is basically an image segmentation problem, we can employ existing DL-based medical image segmentation architectures, such as UNet (Ronneberger et al., 2015), variants of UNet (Alom et al., 2018; Chen et al., 2021; H. Huang et al., 2020; Jing et al., 2022; Oktay et al., 2018; Valanarasu & Patel, 2022; Yang. Y, Zheng. Z, Yuan. Y, Lei. X, 2019; Z. Zhou et al., 2020) and etc, for the construction of the DNN of TSCLP. The subsequent key point is to set the learning strategies for S1 and S2. Since the learning architecture is a DNN, based on an appropriate loss function, we utilized an optimizer of SGD variant to optimize the parameters of the learning models for both S1 and S2 via minimizing the errors between the predictions and the targets. In addition, since S1 and S2 are assembled in a from-coarse-to-fine cascade learning paradigm in TSCLP, the learning rate for S1 should be equal to or larger than the learning rate for S2.

#### 3.2.3. Evolvement of CDSNLL

Firstly, feeding LAND as the input of S1 of TSCLP, we optimized the DL-based medical image segmentation architecture of TSCLP to produce the pre-trained model using a specific optimizer with a specific learning rate. Then, feeding SANFD as the input of S2 of TSCLP, we continued to optimize the pre-trained model to produce the final predictive model using a specific optimizer with a specific learning rate.

In summary, the pseudo codes for the implementation of ECDSNLL are as follows.

Inputs: LAND, SANFD, DNN, Optimizer (Opt), Loss function (Lf), Learning strategy1 (Ls1), Learning strategy2 (Ls2).

Ptm = S1(Opt, DNN, Lf, LAND, Ls1),

= Opt(Lf(DNN(LAND-tiles), LAND-labels), Ls1),

= optimized DNN.

Pm = S2(Opt, Ptm, Lf, SANFD, Ls2),

=Opt(Lf(Ptm(SANFD-tiles), SANFD-labels), Ls2),

= optimized Ptm.

Output: Pm.

Ptm is short for pre-trained model; Pm is short for predictive model.

### 3.3. Evaluating ECDSNLL for PSRTBC

#### 3.3.1. Metrics

We employed usual metrics for image semantic segmentation evaluation. Let TP (true positive) be the number of pixels correctly predicted to belong to the H. pylori class, FP (false positive) be the number of pixels incorrectly predicted to belong to the H. pylori class, and FN (false negative) be the number of missing pixels predicted to belong to the background class. These metrics are tightly related to the foreground class, i. e., the RTBC in which we are interested the most. Based on TP, FP and FN, we further employed recall and foreground intersection over union (fIoU, IoU of the target class) for overall performance evaluation, as the two metrics are important in evaluation of PSRTBC. We also employed PR and ROC curves for the evaluation of the predictive abilities of models. These metrics are commonly used indicators for evaluating the image segmentation performance of a predictive model (Bhattacharjee et al., 2022; Diao et al., 2022a; Elmannai et al., 2021; Priego-Torres et al., 2020, 2022; Yang et al., 2024). More details can also be found in (Wang et al., 2020).

#### 3.3.2. Strategies

Since ECDSNLL is basically a SNLL approach, we followed the basic setting for evaluating SNLL approaches, which in this paper is comparing the performances of models trained on both LAND and SANFD by ECDSNLL with the performances of models trained only on SANFD by supervised learning (SL), to show the effectiveness of ECDSNLL for PSRTBC. For the implementation of this basic setting, SANFD was split into training (SANFD-Train), validation (SANFD-Val) and testing (SANFD-Test) datasets. We respectively trained models on both LAND and SANFD-Train via ECDSNLL and models only on SANFD-Train via SL. Specifically, in training models on both LAND and SANFD-Train via ECDSNLL, we respectively used 80% data of LAND for training and 20% data of LAND for validation to select the pre-trained model; Then, based on the pre-trained model we trained models for evaluation just as the way of training models only on SANFD-Train via SL. For comparisons of the performances between the two types of models, using five-fold cross validation based on SANFD, we respectively trained models for evaluation on SANFD-Train, selected the best models on SANFD-Val, and evaluated their generalization on SANFD-Test. For each fold of cross validation, the SANFD-Test dataset respectively contained 20% data of SANFD, and the SANFD-Train dataset and the SANFD-Val dataset contained 80% data of SANFD. The ratio of SANFD-Train to SANFD-val is 5 to 1.

#### 3.3.3. Details

Based on the metrics and strategies, we conducted comprehensive experiments using state-of-the-art deep neural networks for medical image segmentation to show the effectiveness of ECDSNLL for PSRTBC. The used deep neural networks include the most commonly used UNet (Ronneberger et al., 2015) and its variants from light weight to complex including mobile UNet (MUNet) (Jing et al., 2022; Yang. Y, Zheng. Z, Yuan. Y, Lei. X, 2019), UNeXt (Valanarasu & Patel, 2022), combination of MUNet and UNeXt (MUNeXt), UNet++(Z. Zhou et al., 2020), UNet3+(H. Huang et al., 2020), attention UNet (AttUNet) (Oktay et al., 2018), recurrent residual convolutional neural network based on UNet (R2UNet) (Alom et al., 2018), and TransUNet (Chen et al., 2021). Specifically, based on cross-entropy loss, we utilized the SGD variant Adam (Kingma & Ba, 2014) as the optimizer with the learning rate set to $10^{-4}$ to update the parameters of these UNet variants respectively for SL and both S1 and S2 of ECDSNLL. When calculating the usual segmentation metrics, we used 0.5 to thresh the logits of optimized deep neural networks for PSRTBC, as it is a default value to separate the predictions into tumor and non-tumor which can balance the bias and variance of the optimized deep neural networks.

### 3.4. Producing predictive model for PSRTBC

Firstly, referring to the results of the experiments conducted in evaluation of ECDSNLL for PSRTBC, we selected the appropriate solution to produce the final predictive model for PSRTBC. Then we employed LAND and 80% data of SANFD for training, 10% data of SANFD to select the final predictive model for PSRTBC, and the rest 10% data of SANFD to test the selected predictive model. This data split is different from the strategy presented in evaluating ECDSNLL for PSRTBC, for the purpose here is to produce the final predictive model with the evaluation of ECDSNLL for PSRTBC.

### 3.5. Assessing under microscope scenario

To test the generalization of the predictive model for PSRTBC produced with samples of whole slide imaging to out-of-distribution data, we further qualitatively assessed it under microscope scenario. For the testing, we collected 200 vision fields using a Nikon microscope. For each vision field, we captured images at both 20X and 10X. During testing, the captured vision fields are respectively down sampled into 5X to adapt to the scale the images for training the predictive model for PSRTBC.

## 4. Result and Discussion

This section is structured as follows. In section 4.1, we give the results for the evaluation of ECDSNLL for PSRTBC. In section 4.2, we give the results for the production of a predictive model for PSRTBC. Ins section 4.3, we provide the results for the assessment of the produced model for PSRTBC under microscope scenario. In section 4.4, we describe the release of the produced model for PSRTBC. Finally, in section 4.5, we respectively discuss the results.

**Table 1**

**Evalution results of various SL-based UNet variants. Bold indicates the best solution**

| DNN(SL) | TP↑ | FP↓ | FN↓ | recall(%)↑ | fIoU(%)↑ |
|---|---|---|---|---|---|
| MUNet | 4460 | 430 | 1005 | 81.66 | 75.69 |
| UNeXt | 4379 | 591 | 1085 | 79.84 | 72.01 |
| MUNeXt | 4436 | 381 | 1029 | 81.15 | 75.88 |
| **UNet** | **4767** | **555** | **697** | **87.30** | **79.26** |
| UNet++ | 4649 | 542 | 816 | 85.04 | 77.41 |
| UNet3+ | 4688 | 624 | 777 | 85.85 | 77.18 |
| AttUNet | 4596 | 541 | 869 | 84.14 | 76.49 |
| R2UNet | 4728 | 535 | 737 | 86.41 | 78.69 |
| TransUNet | 4639 | 759 | 826 | 84.79 | 74.77 |
| lower bound | 4379 | 759 | 1085 | 79.84 | 72.01 |
| mean value | 4594 | 551 | 871 | 84.02 | 76.38 |
| upper bound | 4767 | 381 | 697 | 87.30 | 79.26 |

**Table 2**

**Evaluation results of various ECDSNLL-based UNet variants. Bold indicates the best solution**

| DNN(ECDSNLL) | TP↑ | FP↓ | FN↓ | recall(%)↑ | fIoU(%)↑ |
|---|---|---|---|---|---|
| MUNet | 4496 | 443 | 969 | 82.26 | 76.11 |
| UNeXt | 4676 | 664 | 789 | 85.47 | 76.22 |
| MUNeXt | 4528 | 425 | 936 | 82.83 | 76.87 |
| **UNet** | **4828** | **592** | **637** | **88.35** | **79.73** |
| UNet++ | 4738 | 577 | 726 | 86.67 | 78.45 |
| UNet3+ | 4676 | 595 | 789 | 85.62 | 77.27 |
| AttUNet | 4801 | 598 | 664 | 87.89 | 79.23 |
| R2UNet | 4733 | 531 | 731 | 86.47 | 78.82 |
| TransUNet | 4668 | 534 | 797 | 85.40 | 77.89 |
| lower bound | 4496 | 664 | 969 | 82.26 | 76.11 |
| mean value | 4683 | 551 | 782 | 85.66 | 77.84 |
| upper bound | 4828 | 443 | 637 | 88.35 | 79.73 |

**Table 3**

**Evaluation results of performance improvement of ECDSNLL compared with SL. Bold indicates the best improvement**

| DNN (ECDSNLL-SL) | TP↑ | FP↓ | FN↓ | recall(%)↑ | fIoU(%)↑ |
|---|---|---|---|---|---|
| MUNet | 36 | 13 | -36 | 0.6 | 0.42 |
| **UNeXt** | **297** | **73** | **-296** | **5.63** | **4.21** |
| MUNeXt | 92 | 44 | -93 | 1.68 | 0.99 |
| UNet | 61 | 37 | -60 | 1.05 | 0.47 |
| UNet++ | 89 | 35 | -90 | 1.63 | 1.04 |
| UNet3+ | -12 | -29 | 12 | -0.23 | 0.09 |
| AttUNet | 205 | 57 | -205 | 3.75 | 2.74 |
| R2UNet | 5 | -4 | -5 | 0.06 | 0.13 |
| TransUNet | 29 | -225 | -29 | 0.61 | 3.12 |
| lower bound | 117 | -95 | -116 | 2.42 | 4.1 |
| mean value | 89 | 0 | -89 | 1.64 | 1.47 |
| upper bound | 61 | 62 | -60 | 1.05 | 0.47 |
| p-value | 0.015 | 0.499 | 0.014 | 0.016 | 0.009 |

**Figure 3**

**ECDSNLL compared with SL using PR and ROC curves**

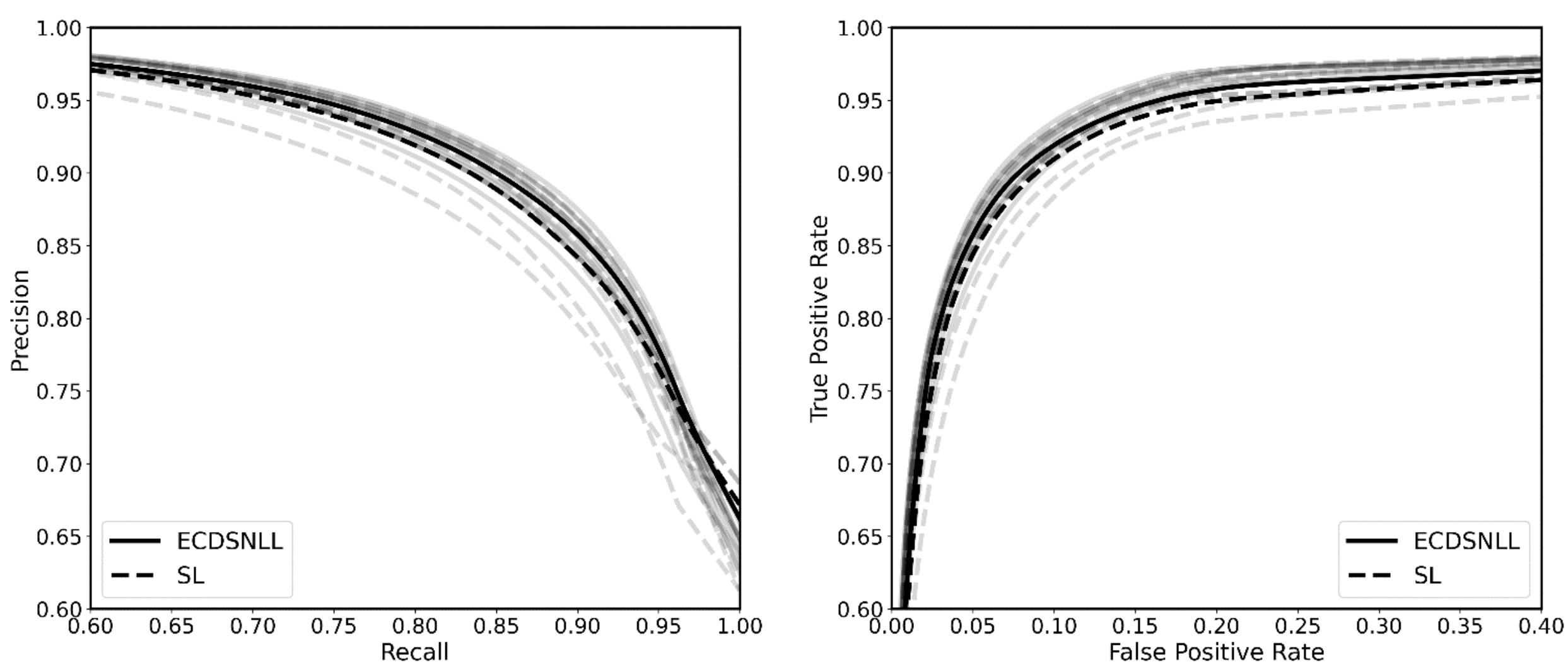

## 4.1. Evaluation result of ECDSNLL for PSRTBC

Following the metrics, strategies and details presented for experiments of evaluating ECDSNLL for PSRTBC in section 3.3, we have following results.

### 4.1.1. ECDSNLL compared with SL using usual segmentation metrics

The evaluation results of various SL-based and ECDSNLL-based UNet variants using usual segmentation metrics are respectively shown as Table 1 and Table 2. And, corresponding to the SL-based and ECDSNLL-based UNet variants in Table 1 and Table 2, the evaluation results of performance improvement of ECDSNLL compared with SL are shown as Table 3.

Based on the employed usual segmentation metrics, we also provide the lower bound, mean value and upper bound of respective metric in Table 1-3, which will be discussed later in the discussion subsection.

#### 4.1.2. ECDSNLL compared with SL using PR and ROC curves

Corresponding to the SL-based and ECDSNLL-based UNet variants in Table 1 and Table 2, the evaluation results of ECDSNLL compared with SL using PR and ROC curves are shown as Fig. 3.

### 4.2. Production result of predictive model for PSRTBC

#### 4.2.1 Predictive model for PSRTBC

Based on the results of the experiments conducted in evaluation of ECDSNLL for PSRTBC and corresponding discussion, we selected the UNet(ECDSNLL) solution to produce the final predictive model for PSRTBC, as it showed the best generalization performance. Following the details described in section 3.4, we produced a UNet(ECDSNLL) solution-based predictive model for PSRTBC.

#### 4.2.2. Quantitative and qualitative evaluation results

For PSRTBC, some quantitative evaluation results of the produced predictive model are shown as Table 4 and Fig. 4, and some qualitative results of the produced predictive model are shown as Fig. 5.

### 4.3. Assessment result of the produced model under microscope scenario

#### 4.3.1 Qualitative results

Some qualitative results of testing the predictive model for PSRTBC produced with samples of whole slide imaging under microscope scenario are shown as Fig. 6 and Fig. 7. Fig. 6 contains the results of vision fields at 20X and Fig. 7 contains the results of the same vision fields at 10X. Particularly, the vision field at 20X and the vision field at 10X were captured at the same vision center for comparisons in control results. The reason that vision fields at 20X and 10X were chosen for qualitative evaluation is because our pathology experts report that they usually do careful inspections under these two magnifications in the microscope scenario and the cases that they do inspections under smaller and larger magnifications are relatively less.

**Table 4**

**Quantitative evaluation results of UNet(ECDSNLL) solution-based predictive model for PSRTBC using usual segmentation metrics**

| Solution | TP | FP | FN | recall(%) | fIoU(%) |
|---|---|---|---|---|---|
| UNet(ECDSNLL) | 5520 | 553 | 669 | 90.89 | 81.88 |

Figure 4

Quantitative evaluation results of UNet(ECDSNLL) solution-based predictive model for PSRTBC using PR and ROC curves

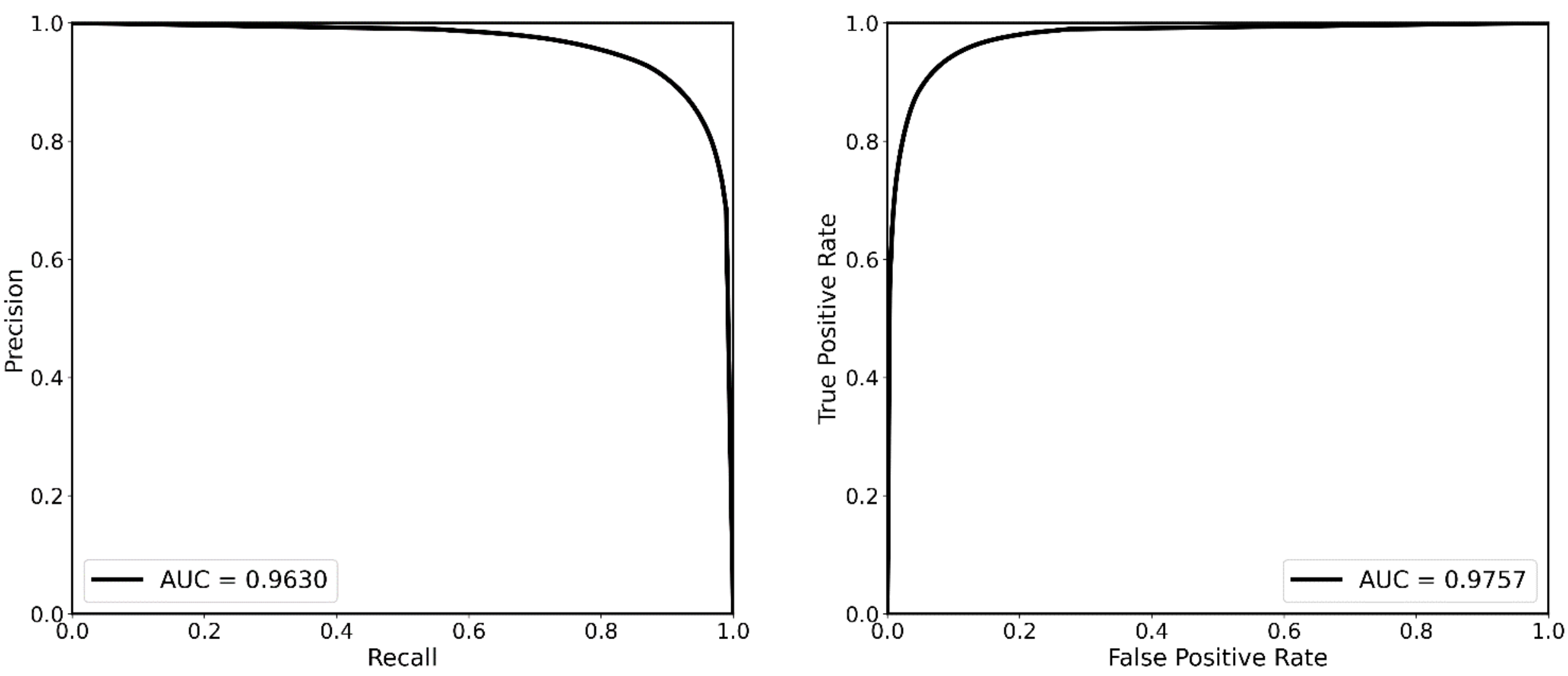

Figure 5

Qualitative results of UNet(ECDSNLL) solution-based predictive model for PSRTBC. Top: input tiles. Middle: manual labels (ground truth) of residual tumor shown on input tiles. Bottom: predicted labels of residual tumor shown on input tiles

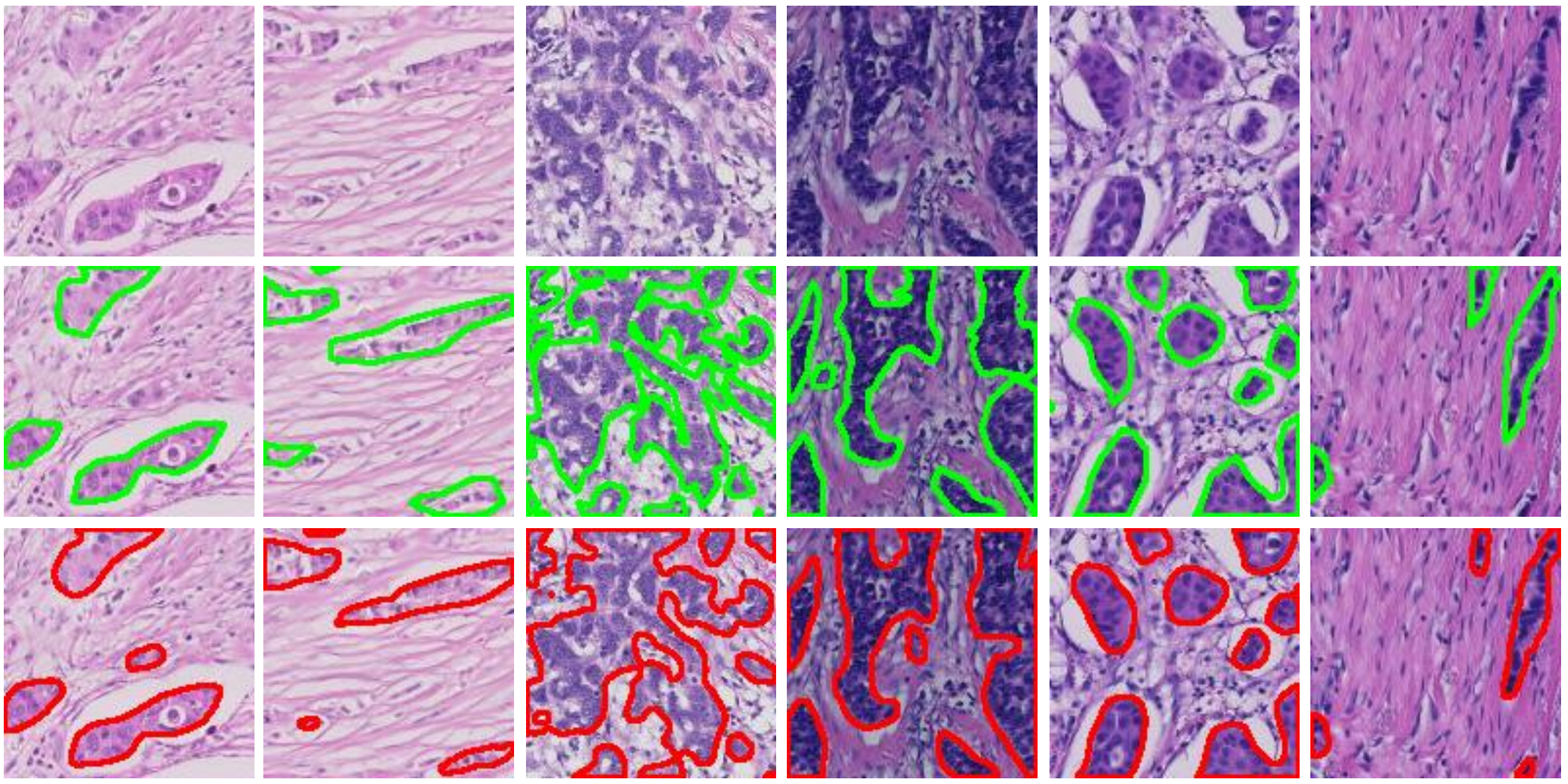

**Figure 6**

**Qualitative results of vision fields captured at 20X. Top row: original images of vision fields; Bottom row: predicted results shown on original images**

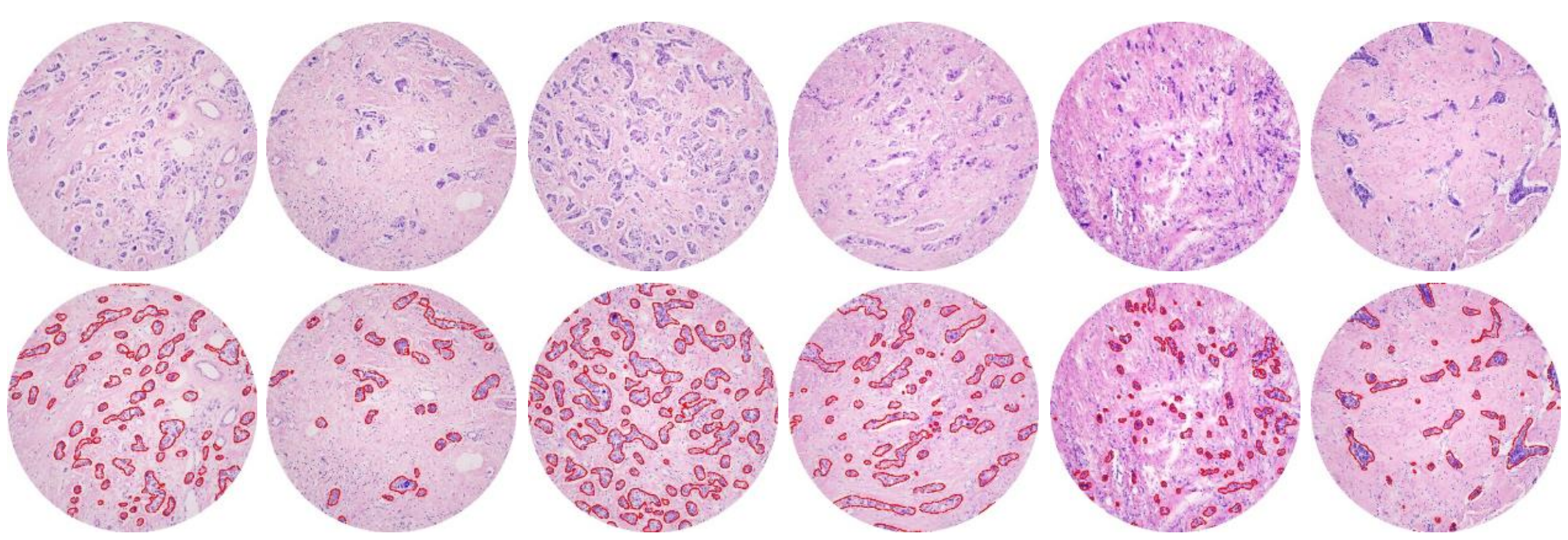

**Figure 7**

**Qualitative results of vision fields captured at 10X. Top row: original images of vision fields; Bottom row: predicted results shown on original images**

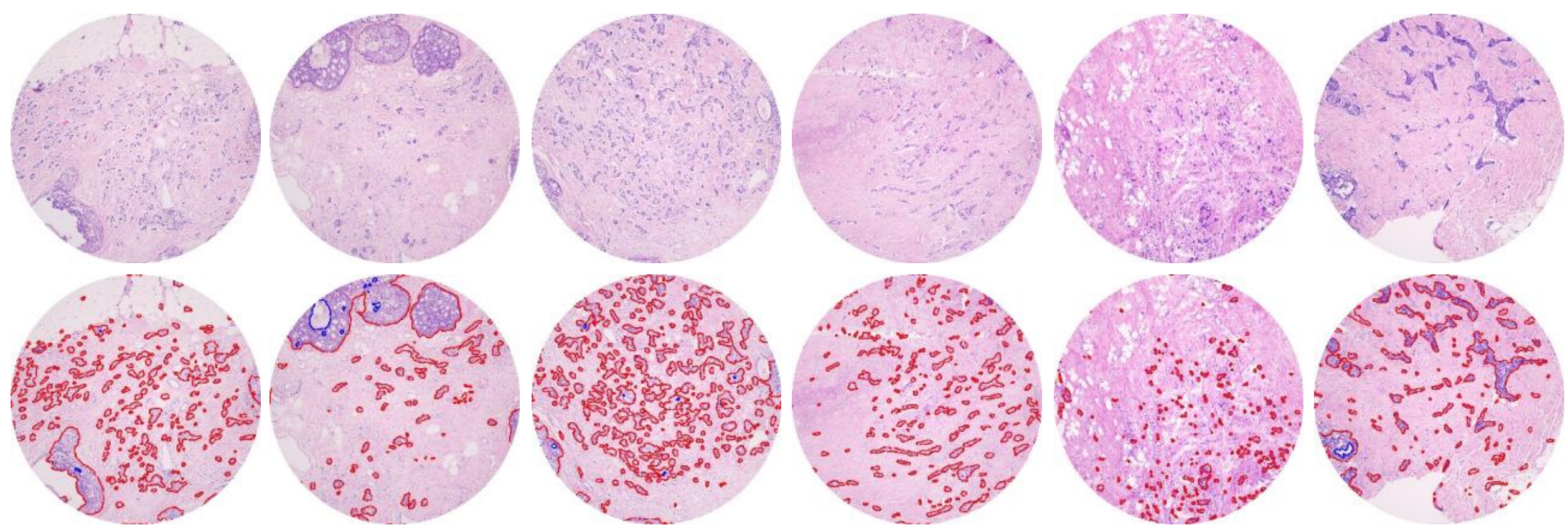

## 4.4. Discussion

From Table 3, we can note that, compared with SL, ECDSNLL can significantly improve the lower bound of a number of UNet variants with 2.42% and 4.1% respectively in recall and fIoU for PSRTBC, while being able to achieve improvements in mean value and upper bound as well. This advantage of ECDSNLL compared with SL is further proved by Fig. 3, as the PR and ROC curves of a number of ECDSNLL-based UNet variant models are respectively covered by the PR and ROC curves of the same number of SL-based UNet variant models, which shows that ECDSNLL-based UNet variant models possess better predictive abilities than SL-based UNet variant models for PSRTBC. Above all, ECDSNLL improves the recall performance while being able to promote the fIoU performance. In addition, statistical test results (p-values) in the last row of Table 3 show that the advantages of ECDSNLL are statically significant compared with SL.

From Table 3, we can also note that ECDSNLL achieves improvements of 5.63% and 4.21% respectively in recall and fIoU with UNeXt. As UNeXt is a very efficient model (Valanarasu & Patel, 2022), these results exhibit the promising potentials of applying ECDSNLL to real-world applications.

Form Table 1, Table 2 and Table 3, we can note that UNet achieves the best performances in both ECDSNLL-based and SL-based solutions, and the UNet(ECDSNLL) solution performs better than the UNet(SL) solution. This indicates that it is appropriate to select the UNet(ECDSNLL) solution to produce the predictive model for PSRTBC.

From Table 4, Fig. 4 and Fig. 5, we can note that the produced UNet(ECDSNLL) solution-based predictive model shows a good performance for PSRTBC in samples of whole slide imaging, which reflect the potentials of applying ECDSNLL in real-world applications.

Two pathological experts were invited to check the qualitative results of the produced predictive model under microscope scenario, and they confirmed that the predictive model for PSRTBC produced with samples of whole slide imaging have the generalization potentials in some vision fields under microscope scenario. At the meantime, they also suggested that the results of vision fields captured at 20X are better than the results of vision fields captured at 10X. This shows the generalization potentials of the UNet(ECDSNLL) solution-based predictive model in some vision fields under microscope scenario for PSRTBC.

## 4. Conclusion and Future Work

Since the breast cancer tissue and tumor cells commonly have complex and varied morphological changes after neoadjuvant chemotherapy (Rajan et al., 2004), achieving precise segmentation of residual tumor in breast cancer (PSRTBC) is still a challenge demanding prompt solution. Although a number of DL-based image semantic segmentation solutions have been proposed for PSTBC of pre-treatment biopsy (Bhattacharjee et al., 2022; Diao et al., 2022a; Elmannai et al., 2021; Priego-Torres et al., 2020, 2022), the fact that breast cancer tissue and tumor cells commonly have complex and varied morphological changes after neoadjuvant chemotherapy (Rajan et al., 2004) makes achieving PSRTBC more difficult than achieving PSTBC.

To alleviate this situation, in this paper, we propose an experts' cognition-driven safe noisy labels learning (ECDSNLL) approach, which integrates the pathology experts' cognition and the artificial intelligence experts' cognition to construct an automatic system for PSRTBC. The proposed ECDSNLL approach has three novel innovations, including making the labeling rules by referring to the pathological experts' cognition about the identification of RTBC to prepare a large amount of noisy data (LAND) and a small amount of noisy-free data (SANFD), establishing the methodology of learning from both LAND and SANFD by referring to the artificial intelligence experts' cognition about achieving better overall performance in prediction to propose a two-stage cascade learning paradigm (TSCLP), and constructing ECDSNLL by appropriately feeding the pathological experts' cognition-driven LAND and SANFD to the artificial intelligence experts' cognition-driven TSCLP. These three innovations eventually make the proposed ECDSNLL approach different from the existing work that realized SNLL in the ensemble learning paradigm, and the related works that realized DL-based predictive models based on separate noisy-free data or separate noisy data.

Compared with the usual supervised learning (SL) paradigm based on SANFD, the proposed ECDSNLL approach based on bothe LAND and SANFD can significantly improve the lower bound performance of a number of UNet variants (Alom et al., 2018; Chen et al., 2021; H. Huang et al., 2020; Jing et al., 2022; Oktay et al., 2018; Ronneberger et al., 2015; Valanarasu & Patel, 2022; Yang. Y, Zheng. Z, Yuan. Y, Lei. X, 2019; Z. Zhou et al., 2020) for PSRTBC, while being able to achieve improvements in the mean value and upper bound performances as well. This advantage of ECDSNLL compared with SL is further proved by the PR and ROC curves of ECDSNLL-based UNet variant models and SL-based UNet variant models, which shows that ECDSNLL-based UNet variant models possess better predictive abilities than SL-based UNet variant models for PSRTBC. The proposed ECDSNLL approach achieved the best performance improvements with a very efficient model UNeXt (Valanarasu & Patel, 2022), which exhibits the promising potentials of applying ECDSNLL to real-world applications. The UNet (Ronneberger et al., 2015) achieved the best performances in both ECDSNLL-based and SL-based solutions for PSRTBC, and the UNet(ECDSNLL) solution performs better than the UNet(SL) solution. The UNet(ECDSNLL) solution was selected to produce the final predictive model for PSRTBC due to its best generalization performance. The produced UNet(ECDSNLL) solution-based predictive model showed a good performance for PSRTBC in samples of whole slide imaging, and had the generalization potentials in some vision fields under microscope scenario.

In this paper, we have shown the advantages of the proposed ECDSNLL approach and its promising potentials in addressing PSRTBC. However, due to the fact that limited number (291) of whole slide images (WSIs) were collected, the produced UNet(ECDSNLL) solution-based predictive model will inevitably have difficulties in handling complicated situations in PSRTBC, where the testing samples of whole sliding imaging are quite different from the samples of our collected WSIs in this paper. Moreover, the produced UNet(ECDSNLL) solution-based predictive model can only be generalized to some vision fields under microscope scenario. As a result, further advances still need to be made to achieve the product-level PSRTBC. However, the predictive model for PSRTBC released in this paper can provide a better foundation for some possible future advances to be made. In addition, it is also interesting to explore additional experiments to test the generalizability of the ECDSNLL approach across different cancer types or imaging modalities (Jain et al., 2023; Murthy & Bethala, 2023), since the concept behind the proposed ECDSNLL approach is fundamentally a new paradigm for artificial intelligence alignment (Yang, 2023) that can be widely leveraged in building predictive models.